\documentstyle[graphics,onecolumn]{mn}
\newcommand{\mincir}{\raise
-2.truept\hbox{\rlap{\hbox{$\sim$}}\raise5.truept 
\hbox{$<$}\ }}
\newcommand{\magcir}{\raise
-2.truept\hbox{\rlap{\hbox{$\sim$}}\raise5.truept
\hbox{$>$}\ }}
\newcommand{\minmag}{\raise-2.truept\hbox{\rlap{\hbox{$<$}}\raise
6.truept\hbox
{$>$}\ }}

\newcommand{\lya}{Lyman-$\alpha$~}
\newcommand{\lyb}{Lyman-$\beta$~}

\newcommand{\vm}{{\rm s/km}}

\newcommand{\be}{\begin{equation}}
\newcommand{\ee}{\end{equation}}
\newcommand{\ba}{\begin{eqnarray}}
\newcommand{\ea}{\end{eqnarray}}
\newcommand{\brr}{\begin{array}}
 
\newcommand{\err}{\end{array}}
\newcommand{\bc}{\begin{center}}
\newcommand{\ec}{\end{center}}

\newcommand{\hm}{\,h^{-1}{\rm Mpc}}
\newcommand{\vel}{\,{\rm km\,s^{-1}}}

\DeclareMathAlphabet{\mathsc}{OT1}{cmr}{m}{sc}
\def\testbx{bx}%
\DeclareRobustCommand{\ion}[2]{%
\relax\ifmmode
\ifx\testbx\f@series
{\mathbf{#1\,\mathsc{#2}}}\else
{\mathrm{#1\,\mathsc{#2}}}\fi
\else\textup{#1\,{\mdseries\textsc{#2}}}%
\fi}

\title[The power spectrum of the flux distribution in 
the  \lya forest]{The power spectrum of the flux distribution in 
the  \lya forest of a Large sample of UVES 
QSO Absorption Spectra (LUQAS)\thanks{
Based on data taken
from the ESO archive obtained with UVES at VLT, Paranal, Chile.}}

\author[T.-S. Kim, M. Viel, M.G. Haehnelt, R.F. Carswell, S. Cristiani] 
{T.-S. Kim $^{1}$, M. Viel $^{1}$, 
M.G. Haehnelt $^{1}$, R.F. Carswell $^{1}$, S. Cristiani $^{2}$
\\ 
$^1$ Institute of Astronomy, Madingley Road, Cambridge CB3 0HA\\
$^2$ INAF-Osservatorio Astronomico di Trieste, via G.B. Tiepolo 11,
I-34131 Trieste, Italy\\
\\}

\begin{document}

\date{submitted 4 April 2003}

\maketitle

\begin{abstract}The flux power spectra of the
\lya forest from a sample of 27 QSOs taken with the high resolution
echelle spectrograph UVES on VLT are presented.  We find a similar
fluctuation amplitude at the peak of the ``3D'' flux power spectrum at $k
\sim 0.03\,{\rm s/km }$ as the study by Croft et al. (2002), in the same
redshift range.  The amplitude of the flux power spectrum increases
with decreasing redshift if corrected for the increase in the mean flux
level as expected if the evolution of the  flux power spectrum is 
sensitive to the gravitational growth of matter density
fluctuations.  This is in agreement with the findings of  
McDonald et al. (2000) at larger redshift.  The logarithmic slope 
of the ``3D" flux power spectrum, $P_F(k)$, at large scales 
$k<0.03 \,{\rm s/km }$, is $1.4\pm0.3$ {\it i.e.} 0.3 
shallower than that found by  Croft et al. 2002  but 
consistent within the errors. 
\end{abstract}


\section{Introduction}

The prominent absorption features blueward of the \lya emission in the
spectra of high-redshift quasars (QSOs) are now generally believed to
arise from smooth density fluctuations of a photoionized warm
intergalactic medium. Support for the hypothesis that such a
fluctuating Gunn-Peterson effect (Gunn \& Peterson 1965) is responsible
for the \lya forest comes from a detailed comparison of analytical
calculations (e.g. Bi \& Davidsen 1997; Viel et al. 2002a; Matarrese \&
Mohayaee 2002) and
numerical simulations (Cen et al. 1994; Zhang, Anninos \& Norman 1995;
Miralda-Escud\'e et al. 1996; Hernquist et al. 1996, Theuns et al. 1998,
2002; Meiksin, Bryan \& Machacek 2001) with observed absorption spectra
(see Rauch 1998 for a review; Kim et al. 2002). The numerical 
simulations have thereby
demonstrated convincingly that the fluctuations in the \lya optical
depth should reflect the density of the dark matter (DM) distribution
on scales larger than a ``filtering'' scale related to the Jeans length
(Gnedin \& Hui 1998; Viel et al. 2002b).  This still rather new
paradigm for the origin of the \lya forest has led to considerable
interest in using QSO absorption spectra to study the dark matter
distribution. Of particular interest is thereby the possibility to
probe the density fluctuation of matter with the
flux power spectrum of QSO absorption lines. The first study of Croft
et al. (1998) was followed by further investigations by Croft et
al. (1999b, 2002; hereafter C99 and C02) and McDonald et al. (2000;
hereafter M00).  These studies compared the observed flux power
spectrum to that obtained from numerically simulated absorption spectra
in order to constrain the slope and amplitude of the linear dark matter
power spectrum for wave numbers in the range $0.002\, \vm 
<k< 0.05\, \vm$. This corresponds  to wavelength of about 
$1\hm$ to $50\hm$ (comoving) at redshift 2--4 and  extends to scales 
considerably  smaller  than those accessible by galaxy
surveys.  At these small scales the matter power spectrum is sensitive
to a possible cut-off expected if the DM were warm dark matter
(Narayanan et al.  2000) and gives constraints on the matter fraction
in neutrinos (Croft et al. 1999a; Elgaroy et al. 2002).  The redshift
range probed is intermediate between that probed by the CMB and galaxy
surveys and allows to investigate the gravitational growth of structure
and possibly the redshift evolution of dark energy (Seljak et al. 2002;
Viel et al. 2003a). C99 inferred an amplitude and slope which was
consistent with a COBE normalized $\Lambda$CDM model with a primordial
scale invariant fluctuation spectrum (Phillips et al. 2001).  However,
M00 and C02, using a larger sample of better quality data, found a
somewhat shallower slope and smaller fluctuation amplitude. The later
data have been claimed to be in (mild) conflict with a 
primordial scale-invariant, CMB-normalized fluctuation spectrum 
and has been used to argue for
a running spectral index (Bennet et al. 2003; Spergel et al. 2003; Verde
et al. 2003).  However, inferring the matter power spectrum from the 
flux power spectrum is a non-trivial matter 
(Zaldarriaga, Hui \& Tegmark, 2001;  Gnedin \& Hamilton 2002). 
In particular,  Zaldarrriaga et al. (2003)  and Seljak et al. (2003) 
have argued that CO2 underestimated the error in the mean flux 
decrement and therefore  underestimated the errors in the slope and 
amplitude of the matter power spectrum. 
There is thus considerable interest in further theoretical study 
and  accurate observational determination  of the \lya forest 
flux power spectrum. We present here the flux power spectra of a  
new large sample of 27 high-resolution absorption spectra taken with the 
high-resolution, high S/N spectrograph on VLT (the LUQAS sample). The
bispectrum of the flux for the LUQAS sample has been investigated in
Viel et al. (2003c).

The sample and the data reduction are presented in
Section 2. Section 3 describes how we calculate the flux power 
spectra. Section 4 contains  our results and in section 5
we give a comparison to previous  published flux power spectra. 
In section 6 we briefly discuss implications of our results for the 
power spectrum of matter density fluctuations and in section 7 we give
our conclusions.  We will assume $\Omega_m=0.3$,
$\Omega_{\Lambda}=0.7$ throughout the paper. 

\section{The Data}

\subsection{Description of the sample} 
\begin{figure*}
\center\resizebox{.65\textwidth}{!}{\includegraphics{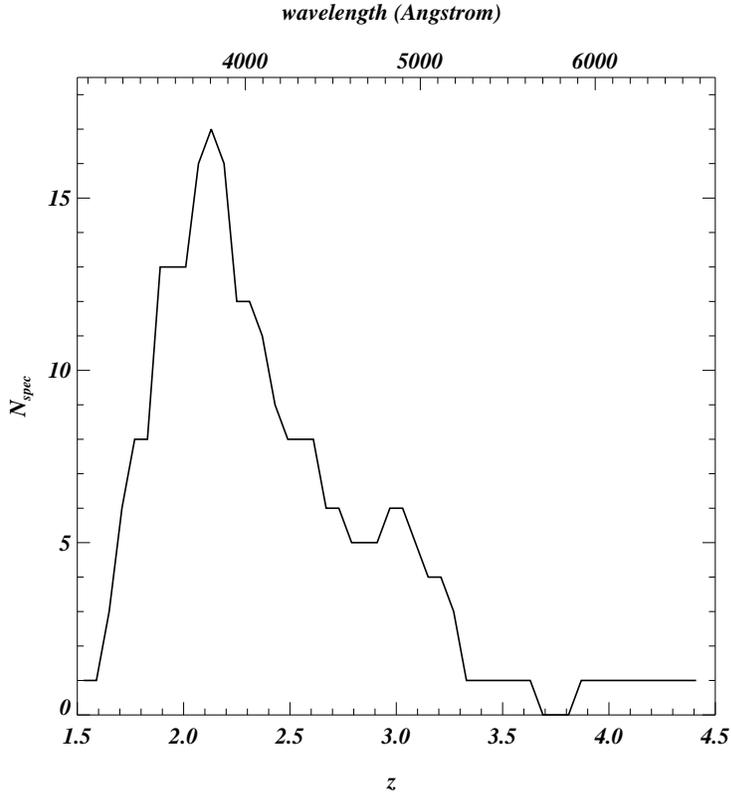}}
\caption{{\protect\footnotesize{Number of spectra  
for which the \lya absorption region covers a  given 
redshift (bottom axis) and wavelength (top axis). The median
redshift of the whole sample is $z=2.25$. The total redshift path is
$\Delta z=13.75$\,.}}}
\label{fig1}
\end{figure*}

The sample consists of 27 spectra taken with the Ultra-Violet Echelle
Spectrograph (UVES) on VLT, Paranal, Chile, over the period 1999--2002.
The 27 spectra were taken from the ESO archive and are publicly
available to the ESO community.  We have selected the QSO sample based
on the following criteria: 1) S/N larger than 25 in the \lya forest
region; 2) complete or nearly complete coverage of the \lya forest
region; 3) few damped \lya systems  and sub-DLAs (column density 
$10^{(19.5-20.3)} \ {\rm cm}^{-2}$) in the \lya forest region; 
4) no broad absorption 
line  systems; 5) publicly available as of January 1, 2003.
Table~\ref{tab1} lists the observation log for the LUQAS sample
including the name and redshift of the targets and program ID. Only
relevant observations are listed.  The total redshift path of the
sample is $\Delta z = 13.75$. In Figure \ref{fig1} we plot the number
of spectra covering a given redshift (bottom axis) and wavelength (top
axis).  The median redshift of the sample is $<z> \, =2.25$ and the
number of spectra covering the median redshift is 17.  In Table
\ref{tab2} we list the \lya redshift range, wavelength range, and
signal-to-noise ratio for all the spectra of our sample.
There is no overlap with published samples for which the 
flux power spectrum has been calculated.

\begin{table*}
\caption[]{Observation Log}
\label{tab1}
\begin{tabular}{lccccccc}
\hline
\noalign{\smallskip}
QSO & $V^{\mathrm{a}}$ & $z_{\mathrm{em}}^{\mathrm{b}}$ &
Setting & Exptime (secs) & Program ID$^{\mathrm{c}}$ 
& UVES/MIDAS$^{\mathrm{d}}$ \\
\noalign{\smallskip}
\hline
\noalign{\smallskip}
HE0001--2340 & 16.7 & 2.259 & dic1 346X580 & 21600 & 160.A-0106 & N\\
             &      &       & dic2 437X860 & 21600 &  & \\
Q0002--422   & 17.5 & 2.659 & dic1 346X580 & 28800 & 160.A-0106 & N\\
             &      &       & dic2 437X860 & 39600 &  & \\
Q0055--269   & 17.9 & 3.656 & dic1 346X580 & 25200 & 65.O-0296 & K2 \\
             &      &       & dic2 437X860 & 21600 &  & \\
Q0103--260   & 18.3 & 3.371 & Red 520      & 24800 & 66.A-0133 & O\\
Q0109--3518  & 16.6 & 2.405 & dic1 346X580 & 25200 & 160.A-0106 & O\\
             &      &       & dic2 437X860 & 21600 &  & \\
Q0122--380   & 17.1 & 2.192 & dic1 346X580 & 21600 & 160.A-0106 & O\\
HE0151--4326 &      & 2.784 & dic1 346X580 & 28800 & 160.A-0106 & N\\
             &      &       & dic2 437X860 & 32400 &  & \\
PKS0237--23  & 16.8 & 2.223 & dic1 346X580 & 21600 & 160.A-0106 & N \\
             &      &       & dic2 437X860 & 21600 & &  \\
Q0302--003   & 18.4 & 3.284 & dic1 346X580 & 20000 & Comm. & K2 \\
PKS0329--255 & 17.1 & 2.707 & dic1 346X580 & 46800 & 160.A-0106 & N \\
             &      &       & dic2 437X860 & 39600 &  & \\
Q0329--385   & 17.2 & 2.435 & dic1 346X580 & 21600 & 160.A-0106 & O\\
             &      &       & dic2 437X860 & 21600 &  & \\
Q0420--388   & 16.9 & 3.117 & dic1 390X564 & 28800 & 160.A-0106 & N \\
             &      &       & dic2 437X860 & 28800 & &  \\
Q0453--423   & 17.3 & 2.657 & dic1 346X580 & 28800 & 160.A-0106 & N\\
             &      &       & dic2 437X860 & 28800 &  & \\
HE0515--4414 & 14.9 & 1.718 & dic1 346X580 & 19000 & Comm. & K1 \\
HE0940--1050 & 16.6 & 3.083 & dic1 346X580 & 18000 & 160.A-0106 \& & N \\
             &      &       & dic2 437X860 & 14400 &  65.O-0474 & \\
Q1101--264   & 16.0 & 2.141 & dic1 346X580 & 23400 & SV & K2 \\
HE1104-1805A  & 17.9 & 2.315 & Blue 360 & 6578  & Comm.  &  O\\
             &       &      & dic1 346X564 & 11400 & 67.A-0278 & \\
             &       &      & dic1 390X580 & 7600 & & \\
             &      &       & dic2 437X860 & 26100 &  & \\ 
HE1122--1648 & 17.7 & 2.405 & dic1 346X580 & 26400 & SV  & K2 \\
             &      &       & dic2 437X860 & 27000 &  & \\
HE1158--1843 & 16.9 & 2.448 & dic1 346X580 & 21600 & 160.A-0106 & N \\
             &      &       & dic2 437X860 & 21600 &  & \\
HE1341--1020 & 17.1 & 2.135 & dic1 346X580 & 32400 & 160.A-0106 & O \\
HE1347--2457 &      & 2.616 & dic1 346X580 & 21044 & SV, 160.A-0106 & K2, N \\
             &      &       & dic2 437X860 & 32400 & &  \\
PKS1448--232 & 17.0 & 2.218 & dic1 346X580 & 28800 & 160.A-0106 & N \\
             &      &       & dic2 437X860 & 21600 &  & \\
Q1451--15    &      & 4.762 & Red 580      & 36000 & 160.A-0106 & N\\
             &      &       & dic2 437X860 & 28800 &  & N\\
PKS2126--158 & 17.3 & 3.280 & dic1 390X564 & 32400 & 160.A-0106 & O \\
             &      &       & dic2 437X860 & 28800 &  & \\
HE2217--2818 & 16.0 & 2.413 & dic1 346X580 & 16200 & Comm. & K1\\
             &      &       & dic1 390X564 & 10800 &  & \\
J2233--606   & 17.5 & 2.251 & dic1 346X580 & 28800 & Comm. & K1 \\
              &      &       & dic2 437X860 & 21600 &  & \\
HE2347--4342 & 16.3 & 2.880 & dic1 346X580 & 21600 & 160.A-0106 & O \\
             &      &       & dic2 437X860 & 28800 &   & \\
\noalign{\smallskip}
\hline
\end{tabular}
\begin{list}{}{}
\item[$^{\mathrm{a}}$] Taken from the Simbad Database. The blank
fields indicate that there is no information of such fields
in the Simbad Database.
\item[$^{\mathrm{b}}$] Estimated from the spectrum.
\item[$^{\mathrm{c}}$] SV: Science Verification in Feb., 2000.
Comm: UVES Commissioning in Sep.-Oct., 1999. Prog ID: 160.A-0106:
The QSO public survey: The cosmic evolution of the intergalactic
medium 
(PI: J. Bergeron). Prog ID: 65.O-0296: A study of the IGM-galaxy
connection at $z \sim 3$ (PI: S. D'Odorico). 
Prog ID: 65.O-0474: Deuterium abundance in the high redshift QSO
system towards HE0940--1050 (PI: P. Molaro). Prog ID: 66.A-0133:
The nature of the \lya forest of Q0103--260 (PI: D. Appenzeller).  
Prog ID: 67.A-0278: Testing the enrichment mechanism of the IGM
through the study of small scale variations in the metal content
of the $z \sim 2$ Ly$\alpha$ forest.
\item[$^{\mathrm{d}}$] K1: Kim et al. (2001), K2: Kim et al. (2002),
O: the 1.2.0 UVES/MIDAS
version, N: the 1.4.0 UVES/MIDAS version
\end{list}
\end{table*}

\begin{table}
\caption{Analyzed QSOs}
\label{tab2}
\begin{tabular}{lccccc}
\hline
\noalign{\smallskip}
QSO & $z_{\mathrm{Ly\alpha}}$ & $\lambda_{\mathrm{Ly\alpha}}(\rm{\AA})$
& S/N per pixel$^{a}$ & metals$^{b}$ & notes$^{c}$\\
\noalign{\smallskip}
\hline
\noalign{\smallskip}
HE0001--2340 & 1.75--2.23 & 3345--3922 & 50--70 &  & DLA at  3873.1 \AA \\
Q0002--422   & 2.09--2.62 & 3755--4404 & 35--65 & & \\
Q0055--269   & 2.93--3.61 & 4778--5603 & 30--75 & fitted&\\
Q0103--260   & 2.72--3.24 & 4522--5156 & $\sim$ 25 & & gap: 5156--5239 \AA\\
Q0109--3518  & 1.87--2.37 & 3495--4098 & 50--70 & &\\
Q0122--380   & 1.69--2.16 & 3276--3841 & 40--70 & fitted &\\
HE0151--4326 & 2.19--2.75 & 3883--4554 & 50--80 & &\\
PKS0237--23  & 1.73--2.19 & 3319--3879 & 55--75 & & AS at 3934.5 \AA\\
Q0302--003   & 2.95--3.24 & 4807--5156 & 30--60 & fitted &\\
PKS0329--255 & 2.13--2.67 & 3808--4461 & 25--30 & & AS at 4513.7 \AA\\
Q0329--385   & 1.90--2.40 & 3525--4131 & 30--55 & fitted & \\
Q0420--388   & 2.48--3.08 & 4230--4955 & 60--100 & & DLA at 4968.5 \AA,
AS at 5013.6 \AA\\
Q0453--423   & 2.09--2.62 & 3753--4401 & 50--70 & & DLA at 4016 \AA \\               
HE0515--4414 & 1.53--1.69 & 3080--3271 & 30--90 & fitted & \\
HE0940--1050 & 2.45--3.04 & 4190--4914 & 60--90 & & \\
Q1101--264   & 1.65--2.11 & 3224--3780 & 30--70 & fitted & sub-DLA at 3450 \AA\\
HE1104--185  & 1.80--2.23 & 3402--3989 & 30--60 & fitted & \\
HE1122--1648 & 1.88--2.37 & 3507--4098 & 35--65 & fitted & AS at 4156.7 \AA\\
HE1158--1843 & 1.91--2.41 & 3544--4150 & 30--50 & &   AS at 4200 \AA \\
HE1341--1020 & 1.66--2.10 & 3227--3773 & 30--55 & & AS at 3826 \AA \\
HE1347--2457 & 2.05--2.57 & 3711--4352& 50--70 & fitted &\\
PKS1448--232 & 1.72--2.18 & 3303--3873 & 30--60 & & \\
Q1451--15    & 3.86--4.70 & 5912--6934 & 35--60 & & \\
PKS2126--158 & 2.61--3.23 & 4392--5151 & 50--100 & fitted & \\
HE2217--2818 & 1.88--2.38 & 3503--4107 & 35--60 & fitted &\\
J2233--606   & 1.74--2.22 & 3337--3912 & 30--50 & fitted & \\
HE2347--4342 & 2.29--2.84 & 4002--4669 & 40--60 & fitted & AS at 4744.3 \AA\\
\noalign{\smallskip}
\hline
\end{tabular}
\begin{list}{}{}
\item[$^{\mathrm{a}}$]
The signal-to-noise ratio varies across the spectrum. The listed S/N is
estimated in regions without strong absorption.
The S/N  usually increases towards longer wavelengths. 
In the case of very high S/N the QSOs usually 
have a prominent Ly$\alpha$ emission line. These values
are meant as  a rough indication of the quality of the spectrum.
\item[$^{\mathrm{b}}$] 
We specify for which QSO metals lines have been fitted 
(see text for the details).
\item[$^{\mathrm{c}}$] 
We identify DLAs (damped \lya systems), AS (associated systems) and
gaps in the spectra only when these systems affect the Ly$\alpha$
forest regions considered in the study.
\end{list}
\end{table}

\subsection{Data reduction} 

The data were reduced with the  ECHELLE/UVES environment of 
the software package MIDAS. Since the release of the first 
version of UVES/MIDAS in  1999, there have been several upgrades. 
Version  1.2.0 was released in September 2001 and the most up-to-date 
version is 1.4.0 which became available with the MIDAS version
released in September 2002. The sample presented here was reduced 
with different versions of UVES/MIDAS as indicated in Table 1.
The 8 spectra marked with ``K'' have been reduced as in Kim et al.
(2002) and the reader is referred to this paper for details. The
spectra marked with  ``O'' and ``N'' are reduced with versions, 1.2.0
and 1.4.0, respectively,  as described in this section.
To ensure that the quality of the reduction is sufficiently homogeneous
for our purposes we have compared the results from different versions 
of the extraction procedure.  The newest version (1.4.0) gives a
slightly better S/N than the older versions. Overall, however, the 
differences are negligible. 

Cosmic rays were flagged using a median filter and the wavelength
calibration was done using the ThAr lamp. 
After the bias and inter-order background are
subtracted, the optimal extraction procedure of
REDUCE/UVES fits the individual  orders of the echelle
spectrum with a Gaussian distribution along the spatial direction.
The sky background is thereby treated as the base of the Gaussian fit.
In this way the sky background is subtracted and an optimal extraction 
of the QSO spectrum is achieved (Kim et al. 2001). This usually works 
except in regions with saturated absorption.  Saturated absorption lines 
are not distinguishable from the sky background in the 2-dimensional
pre-extracted CCD frames. This causes badly fitted Gaussian profiles
for CCD columns containing regions of saturated absorption. 
This leads to an underestimate of the sky background and
results in flux levels larger than zero for saturated lines. The level of
the underestimated sky subtraction varies from frame to frame and
from wavelength to wavelength. It occasionally reaches up to 5\%. 
There is no systematic way to quantify this effect, except 
that the sky background subtraction is worse at short wavelengths, 
in particular at wavelengths shorter than 3400 \AA\/.
We did not try to correct  the zero-level with a fixed offset.

Standard stars, observed during the same nights, were mainly selected from the Kitt
Peak IIDS Standard Star Manual  and their spectra were 
reduced in the same way as the QSO spectra. This was done 
for each night separately to correct for the blaze function
due to the instrument sensitivity. Unfortunately, most of 
the standard stars  
(as the QSOs) do not have flux calibrated points below 
3300 \AA\/.
Consequently we had to extrapolate the flux calibration below 3300 \AA\/.
We also found many standard stars to have many 
absorption lines at this very high resolution. 
With calibrated data points sampled at every
50 \AA\/ to 100 \AA\/ (larger than a typical one echelle order
$\sim 40$ \AA\/), it is not always possible to correct 
for the blaze function. For the QSOs for which observations 
were spread over a period of one year problems occurred 
with the correction  for the shape of the blaze function  
below 3300 \AA\/. Here the flat-field lamp starts to show a 
non-flat feature and the detector sensitivity decreases very rapidly. 
The flux distributions of the standard stars and the QSOs are also 
quite different. These problems result
in a smoothly varying feature in the continuum 
of the merged spectrum. This varying ``continuum feature" can,
however,  be removed by careful continuum fitting.

Where possible, each order of the extracted spectrum was corrected for
the blaze function.  The corrected order was then cut at both ends
where the flat field errors are rapidly increasing. Finally the orders
were merged with weights chosen to maximize the resultant
signal-to-noise ratio.  QSOs for which the blaze function
correction  was not possible suffer from larger continuum
fitting errors below $\sim 3400$ \AA\/ (corresponding to $z_{Ly\alpha}
\le 1.8$).  The LUQAS sample, however, covers mainly redshifts larger
than 1.8.

In the final merged spectrum the individual extractions are weighted 
corresponding to their S/N and resampled with
0.05 \AA\/ binning. The resolving power is $\sim 45\,000$ 
in the regions of interest.
The wavelengths in the final spectra
are vacuum heliocentric. The S/N generally  increases 
towards longer wavelengths. Table~\ref{tab2}
lists the typical S/N of each QSO spectrum and the 
\lya wavelength and redshift ranges.

\subsection{Continuum fitting} 
\label{contdata}

Unfortunately, there is no objective  way to fit the continuum.
The spectra were broken into  parts. For these a continuum 
was fitted with  5th and 7th order polynomials by choosing  
regions devoid of absorptions (for more details, see Kim et al. 
(2001, 2002)). When we split the spectrum into sections for the 
continuum fitting occasionally there is a mismatch in the overlap 
regions of the continuum fit. The overlap regions of echelle 
orders can also be problematic. In these cases we 
adjust the continuum manually to achieve a smooth continuum 
which is still an acceptable fit to the data. At $z\sim 2.5$
 there are about 3-5 such problematic  regions in the	\lya
forest in each object.  

As we will discuss in more detail in Section 3.4.  continuum fitting is
the limiting factor in using flux power spectra to constrain
fluctuations of the matter density on large scales. 
Note also that due to the imperfect blaze function correction 
of  the data for echelle spectra the quality of  the continuum 
fit will  never be very good over  scales larger than one echelle order, 
typically 30--40 \AA\/ for UVES spectra. 
Continuum fitting is no doubt somewhat arbitrary. 
We have therefore also employed a simple averaging method 
originally suggested by C02. 
This method is a modification of the trend removal technique 
proposed by Hui et al. (2001). In section \ref{conteff} 
we will use spectra processed in this way to  demonstrate some 
effects of the continuum fitting  on the flux power spectrum.

\subsection{Damped \lya systems, associated systems and metal lines}  
\label{metaldata}

Here we are interested only in the \lya forest and only consider the
wavelength range between the Ly$\alpha$ and Ly$\beta$ emission lines of
the QSO. To avoid effects due to the enhanced ionizing flux close to
the QSO (proximity effect), we  consider only the Ly$\alpha$ forest
region more than $3000$ km s$^{-1}$ shortward of the Ly$\alpha$
emission.

There are a number of ``contaminants" which may contribute to the  
flux power spectrum on the scales investigated here, most notably DLAs,
associated systems (AS) and metal line systems. 
DLAs and sub-DLAs are easily identified in the
spectra. We report in Table \ref{tab2} the central wavelength of the DLAs
found in the \lya regions out of the 27 spectra. We remove a region
of 100 \AA\/ centered on each damped \lya system  when we calculate the flux power 
spectrum.

Most QSOs in Table \ref{tab1}  show associated systems, {\it
i.e.}  absorption systems at redshifts very close or higher than the QSO 
emission redshift.  Very strong ASs have been found
in 6 spectra, which show non-negligible Ly$\beta$ absorption lines
at wavelength shorter than the QSO \lyb emission line. 
For these QSOs we increase the minimum wavelength of  the 
\lya range used to exclude the Ly$\beta$ absorption  due to 
ASs.

Full absorption line lists were available for 8 QSOs of the sample  
(marked as ``K" in Table \ref{tab1}; Kim et al. 2001, 2002) 
and we have fitted the metal lines 
of five more QSOS using VPFIT 
(Carswell et al.: http://www.ast.cam.ac.uk/$\sim$rfc/vpfit.html).
We expect that less than  5\% of all metal lines 
(usually very narrow lines)  remained unidentified, and that the
residual contamination of the pixel sample by unidentified metal lines
is smaller than 0.3 \%.
In section 3.4 we will use the 13 QSOs for which the metal lines were
fitted to estimate the metal line contribution to the flux power spectrum.

\section{Calculating the flux power spectra} 
\label{conteff}
\begin{figure*}
\center\resizebox{.65\textwidth}{!}{\includegraphics{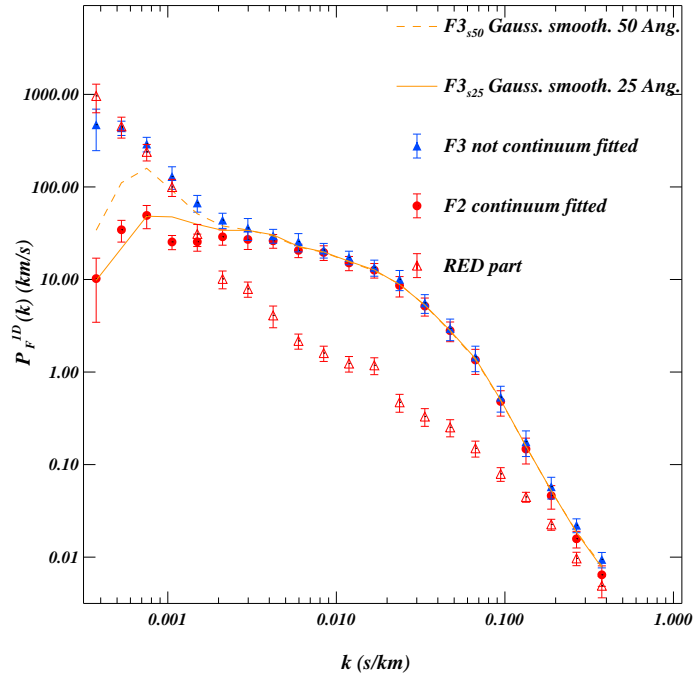}}
\caption{{\protect\footnotesize{
Effect of continuum fitting on the 1D flux power spectrum.  
Circles are for the continuum fitted spectrum $F2=
e^{-\tau}/<e^{-\tau}>-1$, solid 
triangles are for the not continuum fitted spectrum  
$F3 = I_{\rm obs}/<I_{\rm obs}>-1$. Dashed and solid line are for spectra
smoothed with a Gaussian window of 25 and 50 \AA\, width, respectively,
as described in the text. Error bars denote the $1\sigma$ errors of 
the mean values. Open triangles show the flux power spectrum of 
$F3$  for the region  redwards of the \lya emission line 
(1265.67 - 1393.67 \AA). }}}
\label{fig2}
\end{figure*}

\subsection{Basic formulae}  
\label{basic}
The observed intensity is related to the emitted intensity 
through the optical depth, $\tau$, as follows 
\begin{equation} 
I_{\rm obs} = I_{\rm em} e^{-\tau}.
\label{observedI}
\end{equation}
The fluctuations in the observed intensity are thus a superposition 
of the fluctuations of the emitted intensity and those of the
absorption optical depth which reflects the density 
fluctuations of the underlying matter distribution.  
As we will discuss in more detail below                  
continuum fluctuations  start to contribute 
significantly on scales $k< k_{\rm cont} \sim 0.003\, \vm$
and dominate at  $k< \sim 0.001\, \vm$. 
The amplitude of the flux fluctuations will depend on the absolute
flux level. Some sort of  normalization is thus necessary if 
we want to use the flux power spectrum as measure of the 
fluctuations of the optical depth. 

There are several ways to do that. 
Traditionally for a spectrum with high resolution and high S/N, such
as our data, the continuum is estimated by interpolating regions of low 
absorptions with a stiff polynomial (see section \ref{contdata}).
This is a reasonable thing to do
as the  continuum fluctuations dominate only  on large scales. The
power spectrum  of the quantity  
\begin{equation} 
F1= I_{\rm obs}/I_{\rm em}=e^{-\tau}
\end{equation} 
can then be calculated. 
However, the fluctuation amplitude will still depend on the mean 
optical depth in the spectrum,  which is a strong function of
redshift. Because of this Hui et al. (2001) suggested using  the flux power
spectrum of the quantity:   
\begin{equation} 
F2 = e^{-\tau}/<e^{-\tau}>-1 = F1/<e^{-\tau}>-1 \;.
\label{f2}
\end{equation} 
The power spectrum 
of $F2$ is related to that of  $F1$ by a constant factor   
$(<e^{-\tau}>)^{-2}$. 

Continuum fitting is a somewhat arbitrary 
and time-consuming procedure. We have thus also investigated 
the power spectrum of  
\begin{equation} 
F3 = I_{\rm obs}/<I_{\rm obs}>-1= I_{\rm em}e^{-\tau}  /<I_{\rm em}e^{-\tau}>-1, 
\label{f3}
\end{equation} 
where $<I_{\rm em}e^{-\tau}>$ is the average flux of  
the region of the spectrum used to calculate the power spectrum. 
$F3$ is easier to calculate  and its flux power spectrum should become
identical to that of $F3$ at small scales ($k>k_{\rm cont})$
(see also C02). 

To combine and compare spectra at different
redshift it is convenient to introduce yet another estimate of the 
flux 
\begin{equation}
F4=e^{-\tau/B(z)},
\end{equation}
where  $B(z)$ is a scaling factor which varies with redshift and is chosen
in order to reproduce the mean flux at a given $z$.  We will use $F4$
in section \ref{redshiftevolution} to test the gravitational growth of
\lya forest structures.
 When the optical depth is in the range 
$0.05 \le \tau \le 5$, the observed optical depth can be reasonably well 
estimated from the normalized flux (Nusser \& Haehnelt 1999). 
When $\tau$ is smaller the continuum fitting errors and noise 
result in an unreliable estimate of the optical depth. When lines are 
saturated, i.e. $\tau >5$,  the optical depth can 
in principle  be estimated from higher order lines of 
the Lyman series but this is beyond the scope of this paper 
(see for example Aguirre et al. 2002).  In saturated regions we    
leave the flux unchanged. Because of these problem in saturated 
regions and regions of low optical depth this will not 
work too well and F4 should be considered with  caution.  

In order to compute the 1D flux power spectrum, $P_F^{1D}(k)$, we use
the following procedure: {\it i)} we cut the spectrum using the
wavelength ranges indicated in Table \ref{tab2}; 
{\it ii)} we convert the wavelenghts in velocity units assuming an
Einstein-de Sitter Universe (e.g. M00); {\it iii)} we
use a  Lomb periodogram routine (Lomb 1976; Scargle 1982) to compute the
power spectrum of unevenly sampled data.

For an isotropic distribution  (Kaiser \& Peacock 1991) 
the  3D power spectrum is obtained via numerical 
differentiation of the 1D  power spectrum, 
\be
P^{3D}(k)= -\frac{2\pi}{k}\frac{d}{dk} P^{1D}(k).
\label{eq6}
\ee
Both peculiar velocities and thermal broadening make the 
flux field anisotropic and the 3D flux power spectrum 
will  depend on the angle  of the wavevector  to the 
line-of-sight (Kaiser 1987, Hui 1999; McDonald \& Miralda-Escud\'e 1999, 
McDonald 2003). Nevertheless $P_F^{3D}$  calculated 
with eq. (\ref{eq6})  from the 1D flux power spectrum 
is often referred to as 3D flux power spectrum. 
To remind the reader that eq.  (\ref{eq6}) does not give the true 
3D power spectrum we  will denote it  as ``3D'' flux power 
spectrum  throughout.  

To allow a better comparison with previous work 
we will prefer to plot 
the dimensionless  ``3D" power spectrum,
\be
\Delta_F^2(k)=\frac{1}{2\pi^2}\,k^3\,P_F^{``3D"}(k)\, \;,
\ee 
which is the contribution to the variance of the flux per interval $d\ln k$.

C02 pointed out that the ``3D" power spectrum depends
on the resolution of the QSO spectra. Finite resolution results 
in a steepening of the 1D power spectrum. This results in an 
increase of the amplitude of the ``3D" flux power spectrum which 
is obtained via differentiation.

\subsection{Jackknife analysis of statistical errors}  
\label{jackknife}

We use a jackknife estimate for the errors (Bradley
1982). The jackknife estimate of the $1\sigma$ uncertainty of  
a statistical quantity  $Y$ is obtained by dividing the sample 
into $N$ subsamples and computing $\sigma=\left[\sum_{i=1}^N 
(Y_i-\overline{Y})^2\right]^{1/2}$, where $\overline{Y}$ is 
the estimate from the full data sample and $Y_i$ is
the value estimated from the sample {\it without} the subsample $i$.
There are two advantages in using the jackknife estimate
rather than the more common bootstrap resampling techniques.  Firstly, 
it can be 
obtained without any ``randomization'' of the sample. 
Secondly,  it has been shown that jackknife estimates tend to
reduce  the bias which is of order $1/N$ where $N$ is the number of 
measurements. Nevertheless, when we compared the two methods 
they give very similar results. In some cases, especially
for  small subsamples, we will show the 1$\sigma$ dispersion of the  
distribution instead of the jackknife estimate. Note that 
jackknife estimates tend to underestimate the true errors.

\subsection{Effects of continuum fitting and errors 
in the zero level of the flux}

As apparent from eq. (\ref{observedI}) the fluctuations of the flux
distribution are a superposition of variations in the intervening
absorption and variations in the intrinsic spectrum of the QSO  ({\it i.e.}
the continuum).  As
discussed in detail section \ref{contdata} it is possible to remove
most of these fluctuations with a ``continuum fit'' of the
observed spectrum. However, for scales larger than $30-40$ \AA \, 
(this is the typical length of one echelle order) the
continuum fluctuations start to strongly dominate and fundamentally
limit the possibility of recovering the contribution from the intervening
absorption (Hui et al. 2001). This is shown in Figure \ref{fig2} where
we have compared the flux power spectra for continuum fitted spectra
($F2$, eq. \ref{f2}) and spectra without continuum fit ($F3$,
eq. \ref{f3}). As is clearly seen 
the               
continuum fluctuations  start to contribute 
significantly on scales $k< k_{\rm cont} \sim 0.003\, \vm$
and dominate at  $k< k_{\rm cont} \sim 0.001\, \vm$ 
while at smaller scale the flux power spectrum
of the simple averaged quantity $F3$ is identical to that of the
flux-normalized spectrum $F2$. The open triangles show the flux power
spectrum of 22 of the 27 spectra for which the spectra covered the
region 1265.67 - 1393.67 \AA \,  redwards of the \lya emission 
line where there is no 
\lya absorption.  There is a very similar rise 
at  $k< 0.003 \, \vm$. 

The continuum fit has the effect of filtering out fluctuation on scales
larger than 30 \AA. To demonstrate this explicitly 
we show the flux power spectrum
for $F3_{s25} = I_{\rm obs}/<I_{\rm obs}>_{s25}-1$ (and $F3_{s50}$) where
$<I_{\rm obs}>_{s25}$ (and $<I_{\rm obs}>_{s50}$) are the observed
intensity smoothed with a Gaussian window of 25\,\AA \, (50\, \AA)
width  ($\sigma$) as  the solid (dashed) curve.  
At $k> 0.003 \,\vm$ continuum fitting
errors should thus play a very minor role.  The time consuming and
somewhat arbitrary procedure of continuum fitting is thus only
necessary if the recovery of optical depth fluctuations on scales
larger than this is attempted.  In the following we will always take  
$F3_{s25}$ if we do not use  continuum fitted spectra.

The conversion of velocity scale to  physical length 
{\footnote{The different units are
related by:$[k/({\rm km/s})^{-1}]=
(1+z)/(\Omega_m(1+z)^3+\Omega_k(1+z)^2+
\Omega_{\Lambda})^{1/2}[k/h\,(\rm{Mpc}^{-1})/100]=(1+z)/3
\,[k/{\AA}^{-1}]/82.23$.}} is model-dependent. 
For our assumed cosmology ($\Omega_m=0.3$, $\Omega_{\Lambda}=0.7$) 
$k = 0.003 \,\vm$ 
corresponds to a wavelength of $\sim$ 25 $h^{-1}$ comoving Mpc at $z=2.5$.
The fact that the continuum fluctuations contribute significantly at
scales larger than 25 $h^{-1}$ comoving Mpc at $z=2.5$ will make it 
difficult to recover the optical depth fluctuations and thus the 
matter density fluctuation with high accuracy on scales larger than
this scale.  This may,  however, be easier  with the cross-correlation 
of the flux of adjacent lines-of-sight as can be obtained from close
QSO pairs. There the continuum fluctuations drop out because the 
continua of different QSOs will not be correlated on the relevant 
scales unless the QSOs are at very similar redshift (Viel et al. 2002a). 

We have also investigated how  the errors in zero level of the flux discussed 
in section 2.2.  may affect the flux power spectrum. For this purpose
we have set the flux to zero whenever the flux was below 0.05.
The resulting flux power spectrum was different by less than 2 percent.

\begin{figure*}
\center\resizebox{.65\textwidth}{!}{\includegraphics{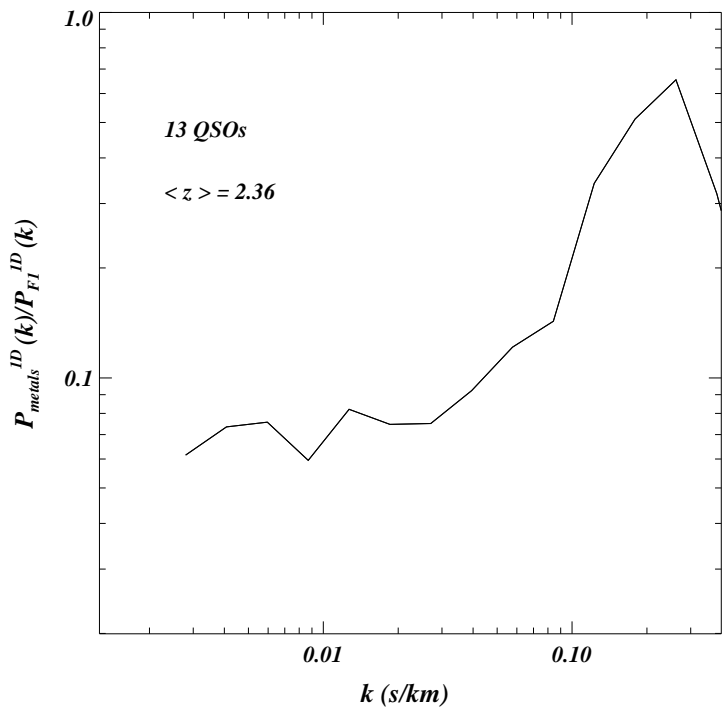}}
\caption{{\protect\footnotesize{Effect of metal lines. The solid curve 
shows the metal line contribution to the total flux power spectrum
of the \lya forest region. Metal-line only spectra have been
generated from the line lists of 13 QSOs fitted with VPFIT. The mean
redshift of this subsample is $<z>=2.36$.}}}
\label{fig3}
\end{figure*}

\subsection{Effects of metal lines and DLAS}  
\label{metaleff}

As discussed in section \ref{metaldata} the \lya forest is
``contaminated'' by metal absorption lines. For 13 of the 
27 spectra (see Table 2) we have identified metal
lines and fitted them with VPFIT. From these line lists
we have produced artificial spectra which contain the metal lines 
only. Figure \ref{fig3} shows the mean flux power spectrum of these metal-line-only 
spectra divided by the mean flux power spectrum of the 
real spectra including metal lines. The contribution of metal lines 
reaches 50 \% at small scales but drops to less than 
10 \% at large scales ($k<0.1 \, \vm$). 
This agrees well with estimates of  M00  
and C02,
who also found the metal line distribution to 
be small on scales larger than $k \sim 0.1 \, \vm$. 

We are here mainly interested in the large scale flux power spectrum 
and thus in the following we will ignore the metal line contribution. 
In principle it is possible to remove the metal lines but this was not
done because it 
can lead to windowing effects in the power spectrum on large scales 
where removal is not really necessary and will be difficult at small
scales because of the 
rather large metal line contribution. Attempts to use the small scale cut-off 
to constrain the matter power spectrum  and to determine physical 
parameters of the \lya forest will need a very careful removal 
of the metal line contribution.  

We have also computed the flux power spectrum with  and without 
cutting out the damped \lya systems. The results are  consistent within
the error bars but generally smaller without damped \lya systems.
The difference is less than 5\% at large scales and  drops to  2\% 
for $k>0.02 \,{\rm s/km }$.  Note that the sample contains 
only four (sub-)damped \lya systems which we have cut out.

\section{Flux power spectrum of the observed absorption spectra} 

\begin{figure*}
\center\resizebox{1.\textwidth}{!}{\includegraphics{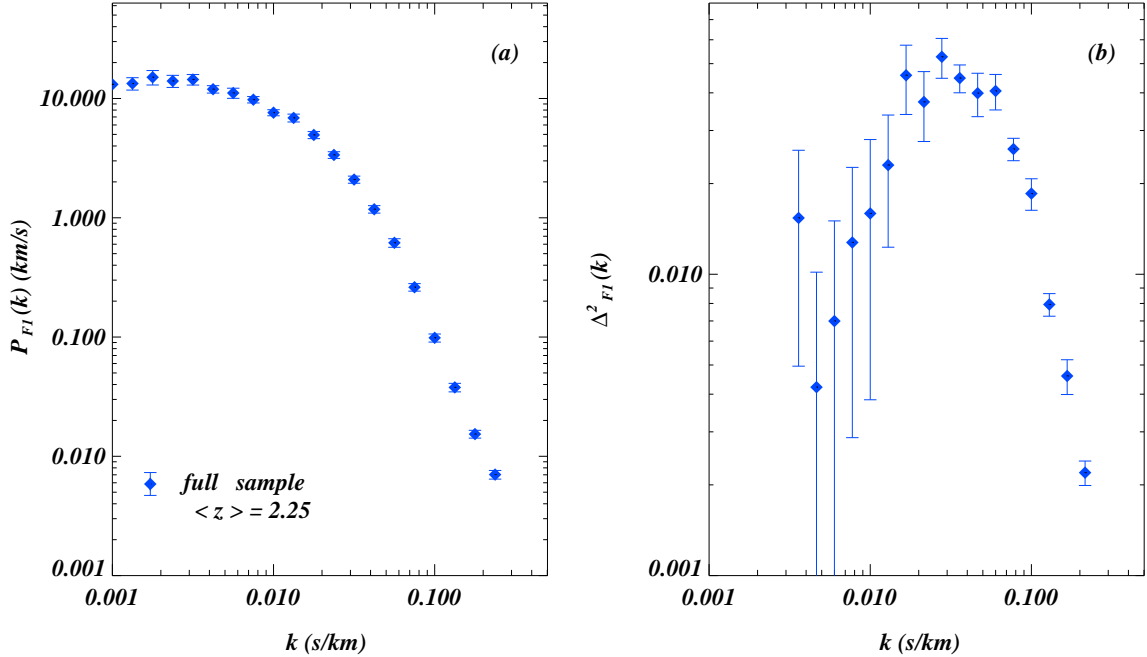}}
\caption{{\protect\footnotesize{The 1D  and ``3D"
flux power spectrum of the full sample for the flux estimator 
$F1=\exp(-\tau)$. Errors bars are jackknife estimates.  Note that the LUQAS points (diamonds) in the left
panel are different from the one in the published version in Kim et
al. (2004) MNRAS 347, 355 (where the $k$ values of the 1D flux power
spectrum had been erroneously shifted by half a bin size in $\log k$).}}}
\label{fig4}
\end{figure*}

\begin{figure*}
\center\resizebox{.65\textwidth}{!}{\includegraphics{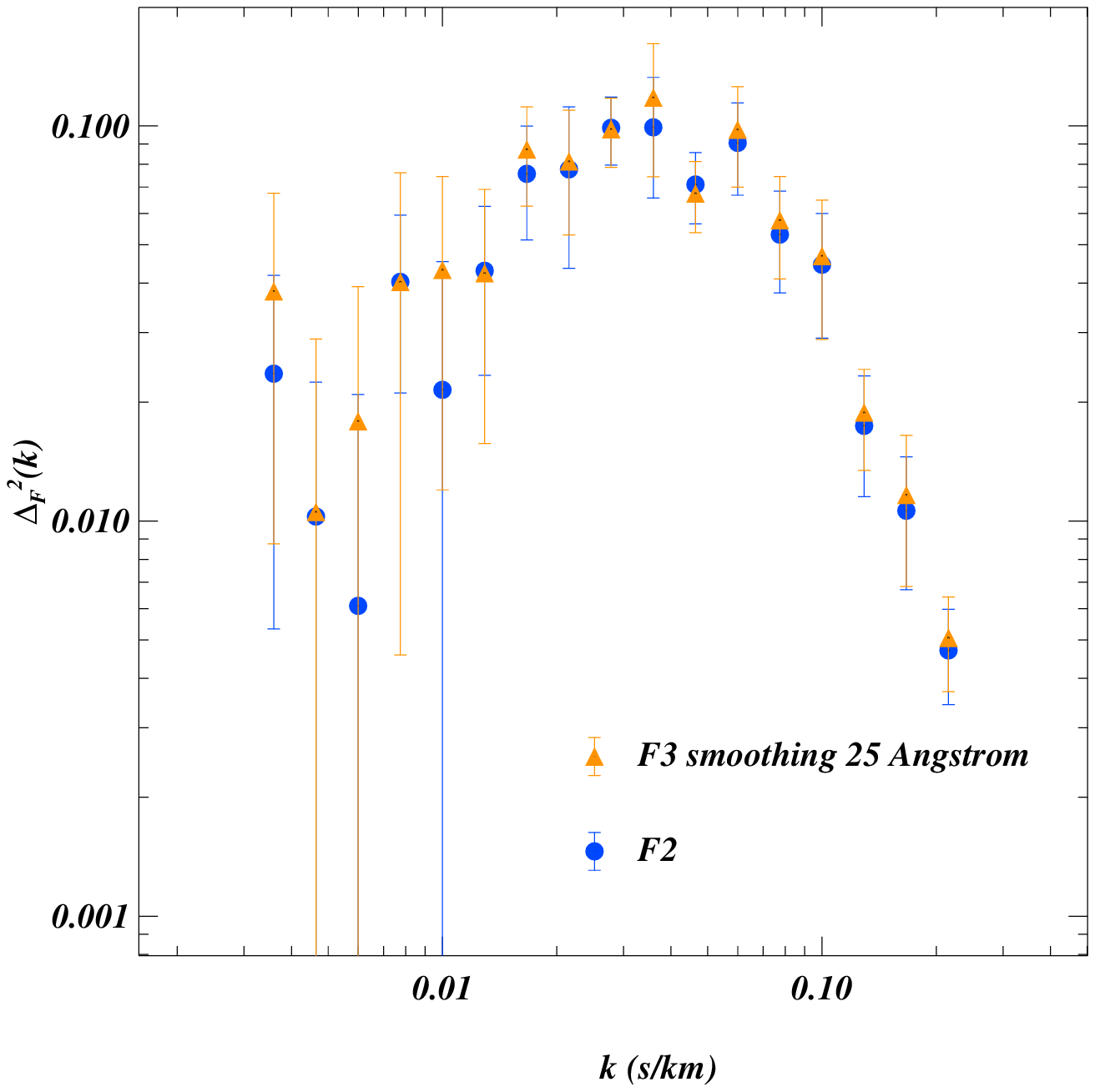}}
\caption{{\protect\footnotesize{``3D" power spectrum of two
different flux estimators. Circles are for 
$F2= e^{-\tau}/<e^{-\tau}>-1$ calculated from continuum fitted spectra
and triangles are for  $F3_{s25} = I_{\rm obs}/<I_{\rm obs}>_{s25}-1$ 
calculated from not continuum fitted spectra. $F3$ is smoothed 
with a Gaussian window of width  25 \, \AA. See Section
\ref{conteff} for details. Error bars are jackknife estimates
calculated from the sample of 27 QSOs.}}}
\label{fig5}
\end{figure*}

\subsection{The total sample} 

In Figure \ref{fig4} we show the 1D and ``3D" power spectrum of the flux
estimator $F1$ for the full sample. The spectra were continuum fitted
as described in section \ref{contdata} and no attempt has been made to
remove metal lines. The 1D spectrum is remarkably smooth for $k > 0.005\,
\vm$. The errors were obtained with the jackknife estimator described
in section \ref{jackknife}.  Figure \ref{fig5} compares the ``3D" power
spectrum of the flux-normalized spectra ($F2$) and those of spectra
smoothed with a $25$ \, \AA \, window. For $ k < 0.007 \, \vm$ the
power spectrum of $F3_{s25}$ is still somewhat larger than that of $F2$
indicating that the smoothing window was too wide to remove the
continuum fluctuations on these scales.  Tables \ref{tab3} and
\ref{tab4} give the mean 1D and ``3D" flux power spectrum of the full
sample for the flux estimator $F1$, $F2$ and $F3$ as described in \ref{basic}
and \ref{conteff}.  The ``3D" flux power spectrum was estimated by
linearly connecting neighbouring points of an estimate of the 1D flux power
spectrum binned with the same bin size but shifted by half a bin
width.  The bin size was chosen such that the visual comparison of different
samples in the plots is not hindered by too large bin to bin
fluctuations and varies for different (sub-)samples.

\begin{figure*}
\center\resizebox{1.0\textwidth}{!}{\includegraphics{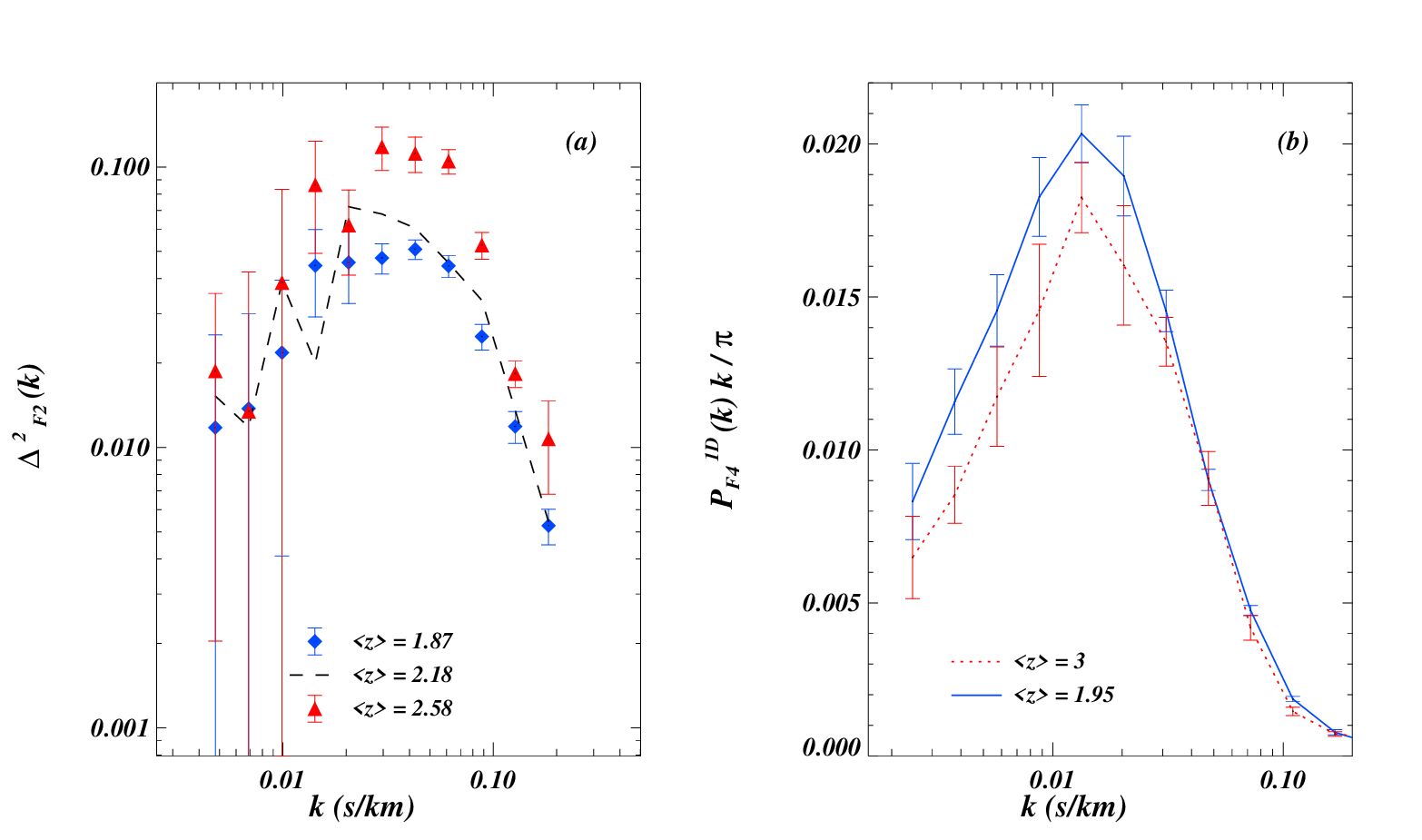}}
\caption{{\protect\footnotesize{{\it Left:} Redshift evolution
of the ``3D" flux power spectrum of the flux estimator 
$F2=e^{-\tau}/<e^{-\tau}>-1$:   diamonds ($1.6 < z < 2$; $<z>=1.87$);
dashed line ($2 < z < 2.4$; $<z>=2.18$); triangles ($2.4 < z < 2.8$; $<z>=2.58$). 
{\it Right:} Redshift evolution of the dimensionless 
1D flux power spectrum  $k P_{F}^{1D}$ after rescaling of the 
optical depth such that $\tau_{eff}=0.15$ at $z=2$ for all spectra 
(flux estimator F4) and for two different mean redshift of $<z>=1.95$
(continuous line) and $<z>=3$ (dotted line). Note that the flux power spectrum now increases
with decreasing redshift as expected for gravitational growth. 
Errors are jackknife estimates.}}}
\label{fig6}
\end{figure*}

\subsection{Redshift evolution} 
\label{redshiftevolution}

The flux power spectrum evolves strongly  with redshift. 
It is thereby important to discriminate between changes due 
to evolution of  the  absolute emitted flux, the normalized flux 
and the optical depth. As discussed in more detail in section 
6 it is the latter which is most directly related to the fluctuations 
of the matter density. Figure \ref{fig6}a shows  the redshift evolution of the 
flux power spectrum for  our standard estimator $F1$. The amplitude  
increases strongly with increasing  redshift.  As discussed 
in detail by C02 this is due to the increased 
mean optical depth at high redshift. 
The amplitude of the matter density fluctuations actually evolves in
the opposite way. Figure \ref{fig6}b shows the evolution of 1D  flux power
spectrum for our estimator $F4$ where we have rescaled the 
optical depth of the observed spectra  to have the same effective 
optical depth of $\tau_{\rm eff}= -\ln <F1> =0.15$,   
the mean value at redshift $z=2$. 
Note that in Figure \ref{fig6}b we have plotted 
$k P_{F}^{1D}$ and that the plot is linear. 
These rescalings are somewhat problematic because both very small 
and very large optical depths cannot be recovered correctly because of 
noise and saturation effects. Nevertheless it is encouraging that 
with the rescaling the evolution is now reversed and the amplitude 
increases with decreasing redshift as expected for the 
fluctuations of the matter density. 
In Tables \ref{tab5} and \ref{tab6} we list the values for the
1D and ``3D" estimates of the flux power spectrum $F2$, in three different
redshift ranges with mean redshift $<z> = 1.87,2.18,2.58$.

\section{Comparison with  previous results} 

\begin{figure*}
\center\resizebox{1.\textwidth}{!}{\includegraphics{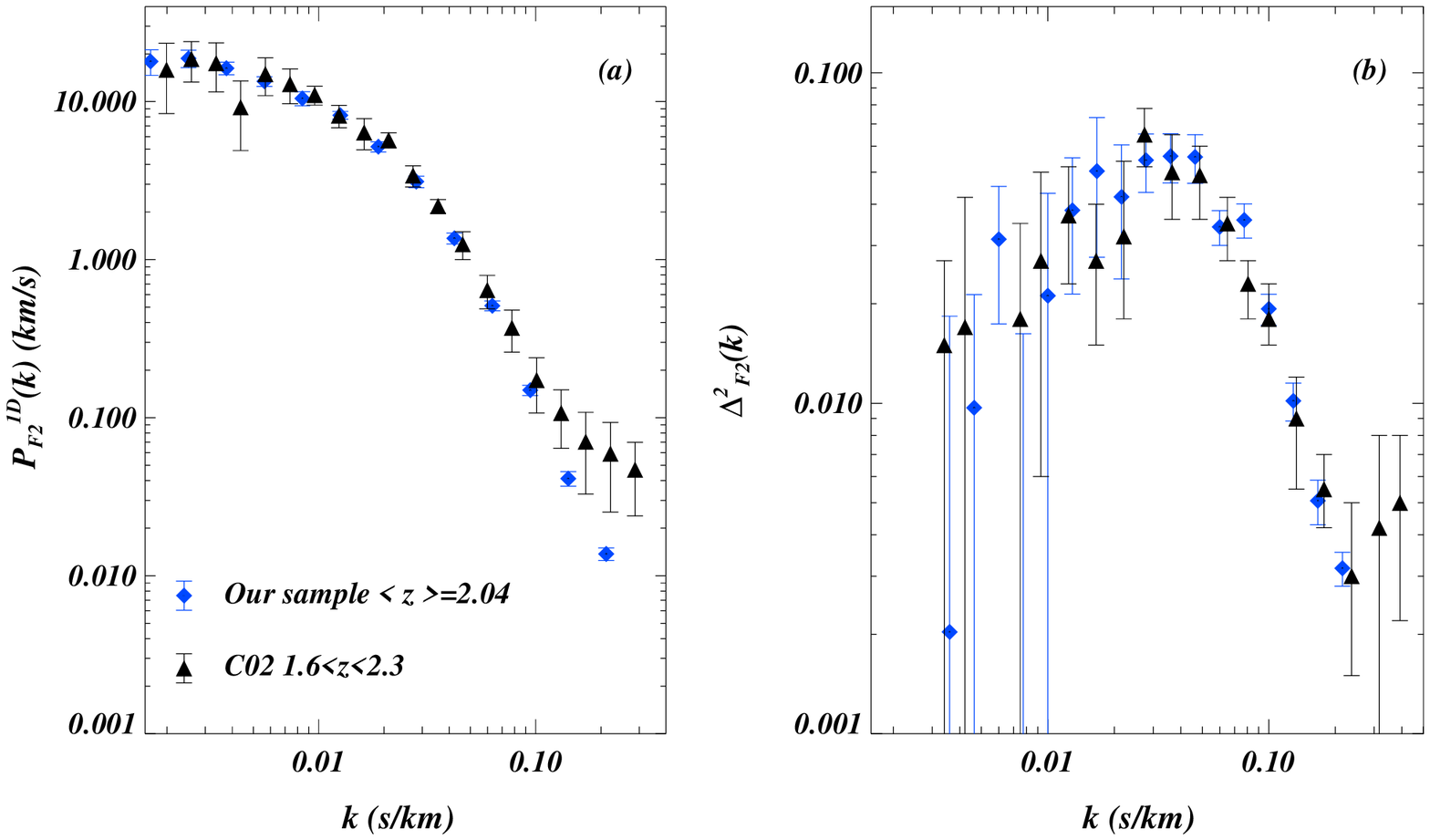}}
\caption{Comparison with Croft et al. 2002 (C02) results.  {\it Left:}
1D power spectrum of $F2= e^{-\tau}/<e^{-\tau}>-1$ for our sample in
the redshift range $1.6<z<2.3$, with a $<z>=2.04$, (diamonds) and the
C02 results for their subsample A in the same redshift range
(triangles).  {\it Right:} ``3D" flux power spectrum (plotted in
dimensionless units) for the two samples on the left. Error bars are
jackknife estimates. Note that the LUQAS points (diamonds) in the left
panel are different from the one in the published version in Kim et
al. (2004) MNRAS 347, 355 (where the $k$ values of the 1D flux power
spectrum had been erroneously shifted by half a bin size in $\log k$).}
\label{fig7}
\end{figure*}

\begin{figure*}
\center\resizebox{1.\textwidth}{!}{\includegraphics{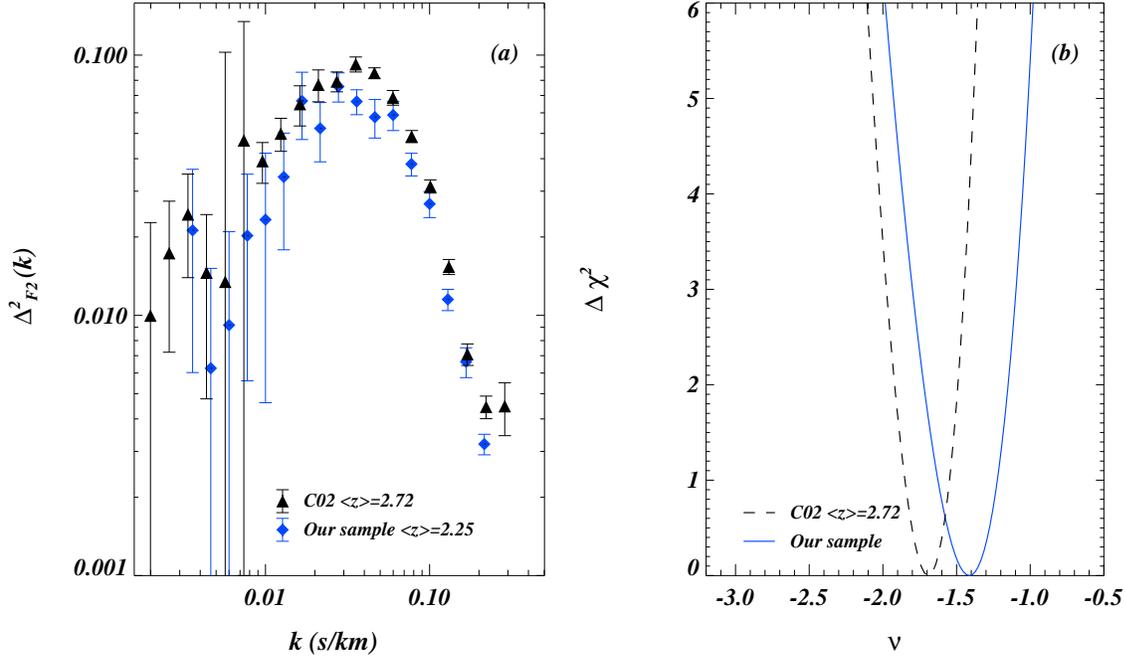}}
\caption{{\it Left:} Comparison of the ``3D" power spectrum 
of our full sample  and that of C02. {\it Right:} $\chi^2$ 
distribution for the logarithmic slope of a power law 
fit to the flux power spectrum, $P_F^{``3D"}(k)$, in the left panel.  
The range of wavenumbers fitted is  $0.0035< k\, (\vm)<0.02$ and 
we fix the amplitude of our fit (eq. \ref{eqfit}) 
to the best fitting value at $k = 0.015 \, \vm$.}
\label{fig8}
\end{figure*}

\begin{figure*}
\center\resizebox{.7\textwidth}{!}{\includegraphics{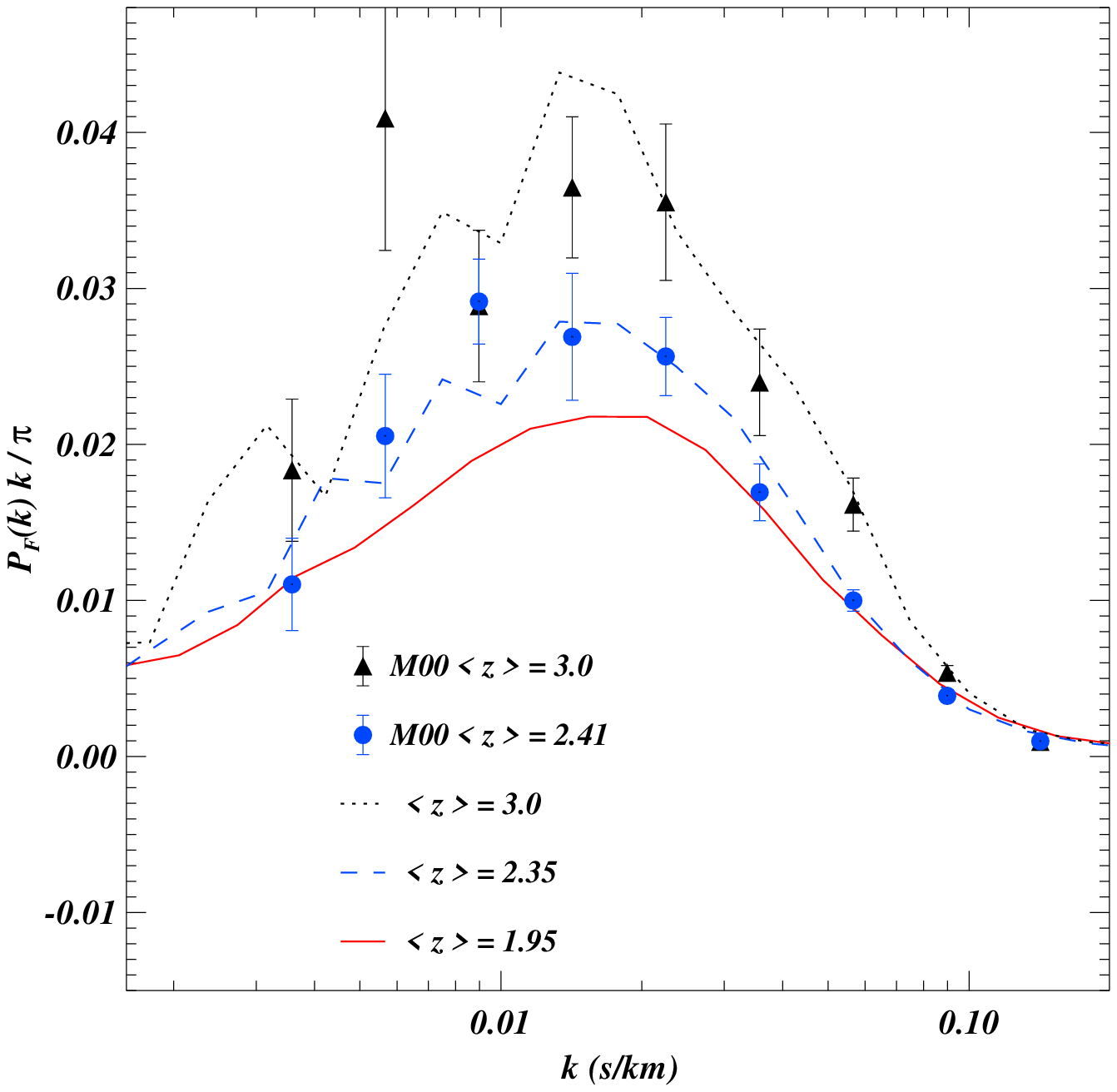}}
\caption{Comparison of redshift evolution to 
McDonald et al. 2000.  Plotted is the dimensionless 1D power spectrum 
$kP_{F}^{1D}$ for three different subsamples of our data set: $1.75<z<2.15$
($<z>=1.95$, continuous line), $2.15<z<2.65$ ($<z>=2.35$, dashed line) and $2.75<z<3.25$
($<z>=3$, dotted line).  The results of McDonald et al. 2000 (M00) at 
two different redshifts 
are shown as circles ($z=2.41$)  and 
triangles ($z=3.0$).}
\label{fig9}
\end{figure*}

\subsection{Amplitude and slope}

We first compare the subsample of our spectra in the redshift range
$1.6<z<2.3$ with subsample A of C02 where we have the biggest overlap
in terms of number of QSO spectra. The 1D flux power spectra for the
flux estimator $F2$ are in good agreement, within the errors, in the range of $k$ not
affected by differences in resolution and S/N (Figure \ref{fig7}a).
The power spectrum of the LUQAS sample is somewhat smoother due to the
larger number of QSO spectra in this redshift range.  There is a slight
``excess'' of our sample at $k\sim 0.04 \,\vm$ compared to the
C02 sample.  At the smallest scales the pixel noise in the
spectra starts to dominate and at $k > 0.3 \,\vm$ the effect of
the about a factor two larger pixel noise in the Croft sample can be
seen.  Hui et al. (2001) gave a rough estimate of the pixel noise
contribution to the flux power spectrum $\sim 2 (\delta v / \bar
f)(S/N)^{-2}$, where $ \delta v$ is the pixel size, $\bar f$ is the mean
transmission and $S/N$ is the signal-to-noise ratio.  For the LUQAS
sample with a pixel size of $2.5$ km/s, $\bar f = 0.86$ and $S/N =50$
the expected contribution to the flux power spectrum is $2.5\times 10^{-3}
\vel $ and thus below the plotted range.
   
In Figure \ref{fig7}b we compare the corresponding ``3D" flux power spectra.  The ``3D"
flux power spectra are also in good agreement and the amplitude is the
same within the errors.  As before the ``3D" power spectrum is considerably
noisier because of the differentiation procedure and we have thus
chosen to compare our full sample to the full sample of C02 (Fig. \ref{fig8}a)
in order to assess whether the LUQAS sample has the same slope at
intermediate scales, where continuum fitting errors are not dominant
and the flux power spectrum is not affected by thermal effects.  The
slope of $P_F^{``3D"}(k)$, at small $k$, for our sample appears to be
somewhat shallower. To quantify this we have fitted a power law of the
form: \be P_F^{``3D"}(k) = P_F^{``3D"}(k_p)\left ( \frac{k}{k_p}
\right)^{\nu} \; ,\label{eqfit} \ee to the ``3D" flux power spectra.

Figure \ref{fig8}b shows the $\chi^2$ distribution for a fit in 
the range $0.0035< k\, (\vm)<0.02$. We have 
thereby followed the procedure in Croft et al. (2002). 
We have first performed  a 2 parameter fit.  
We have then  determined a pivot wavenumber for which 
the amplitude and slope are uncorrelated  
($k = 0.015 \, \vm$ for bout our sample and that of C02) 
and varied the slope  keeping the amplitude at the best fitting 
value of the  2 parameter fit. The resulting $\chi^2$ is shown in 
Figure \ref{fig8}b.  The logarithmic slope of $P_F^{``3D"}(k)$ for
the LUQAS sample is 0.3 shallower  than that of the C02 sample,
although the results are compatible within 1$\sigma$ errors.  Note
also that the median redshift of our sample is somewhat lower than 
that of C02 and that the largest scale used in the fit may already be slightly 
affected by continuum fitting (see Figures \ref{fig2} and \ref{fig5}).

\subsection{Redshift evolution}

In Figure \ref{fig9} we  compare the evolution of the dimensionless 1D power spectrum 
for the flux estimator $F1$ of our three subsamples with that of M00. 
The agreement is again good at the redshifts covered by the M00
sample. Note that we follow M00 and plot $k P_{F}^{1D}$ and 
that the plot is linear.  
The observed decrease of the flux power spectrum of the LUQAS  sample which
extends to lower redshift continues. The flux power spectra of the M00
sample are somewhat noisier because of the smaller number of QSO spectra. 
In section 4.3 we had demonstrated that the amplitude of the flux
power spectrum  of our sample increases with decreasing redshift once 
we rescaled the optical depth. Such a growth is expected if the matter
fluctuations grow. C02 found that their recovered linear
matter power spectrum  was consistent with gravitational growth.  

Our attempt to rescale our spectra to the mean optical depth of the M00
did not give reasonable results. This is most likely due to the fact that a
rescaling of the optical depth by large factors does not work 
because recovering the optical depth at 
small and large values is significantly limited by noise/continuum
fitting errors and saturation effects, respectively.  
A direct comparison of rescaled spectra with  the  M00  
results  was thus not possible.

\section{From the flux power spectrum to the matter power spectrum and 
         cosmological parameters}

In the fluctuating Gunn-Peterson approximation the relation between the absorption optical depth 
and matter density in redshift space can be written as 
\begin{equation} 
\tau =  A  \left ( \frac{\rho}{<\rho>} \right ) ^{\alpha} 
\end{equation} 
where $1.6 \la \alpha \la 2$ depend on the temperature density
relation of the gas (see Weinberg et al. 1999 for a review). 
For small fluctuation    
\begin{equation} 
\frac{dF}{<F>} = - A \,\alpha\, \delta,  
\end{equation} 
where $1+\delta=\rho/<\rho>$.
On large scales the  flux power spectrum should thus, up to a constant 
factor, be identical to the matter power spectrum in redshift space. 
Unfortunately even on large scales this is complicated due to peculiar 
velocity effects and 
the presence of non-linear structures which lead to saturated
absorption. Viel et al. (2003b) find that strong discrete absorption 
systems contribute significantly ($\ga 50 \% $) to the flux power
spectrum at large scales.   
The relation between the flux and matter power spectrum 
is  thus more complicated.  
C99, C02 and M00 have used numerical 
simulations to determine the normalization and slope of the real space matter 
power spectrum from the flux power spectrum. 
The samples have different redshift distributions and resolution. 
It is  thus not completely  straightforward to compare the results. 
After suitable rescaling in $k$ and redshift C02 found good agreement
with M00 and a  25 \% smaller fluctuation amplitude and a 0.1
shallower logarithmic slope of the  matter power spectrum than C99. 
Without numerical simulations we are not able to directly infer the 
amplitude and slope of the linear matter power spectrum. 
However, the amplitude of the flux power spectrum is in good
agreement with that of C02 and M00 and the inferred matter fluctuation
amplitude should  be the same.  We have obtained a  0.3 
shallower  logarithmic slope for $P_F^{``3D"}(k)$ but the errors are
large and the results are consistent with that of Croft et al. within 
1 $\sigma$.    
The recent combined analysis  of the WMAP results with the
C02  has claimed that shape and amplitude of the matter power spectrum 
inferred by C02 is not consistent with a scale invariant initial
fluctuation  spectrum  (Bennet et al. 2003, Spergel et al. 2003, 
Verde et al. 2003). However,  as discussed in the introduction, it 
has been argued that the errors of amplitude and slope 
of the matter power spectrum have been underestimated 
and that this discrepancy is not significant 
(Zaldarriaga, Soccimarro \& Hui, 2003; Zaldarriaga, Hui \& Tegmark, 2001; 
Gnedin \& Hamilton 2002; Seljak , McDonald \& Makarov 2003).

\section{Conclusions}

We have calculated the flux power spectrum for a new large sample of
high resolution high signal-to-noise QSO absorption spectra taken 
with the ESO-UVES spectrograph and compared it to previous published
flux power spectra.  Our conclusions can be summarized as follows.

1.  The flux variation $\Delta_F^2$  rises from 0.01 at $k \sim 0.003
    \,\vm$ to 0.1  at  $ k \sim 0.03 \,\vm$  and drops 
    rapidly at larger $k$.

2. At $k < 0.003 \,\vm$    the flux power spectrum is 
   strongly dominated by  continuum fluctuations while at 
   $k > 0.003 \,\vm$ the  results for continuum fitted spectra 
   and simply averaged spectra are very similar.  
 
3. The metal line contribution to the flux power spectrum is 
   less than 10 \% at scales  $k < 0.01 \,\vm$ 
   and rises to 50 \% at smaller scales. 

4. The agreement with the studies of Croft et al. (2002 - C02) and
   McDonald et al. (2000 - M00) is good. The amplitude of the flux
   power spectrum at the peak is the same as that in the sample of
   C02 and M00 within the errors.  The logarithmic slope for the flux
   power spectrum $P_F^{``3D"}(k)$ at small $k$ is 0.3 shallower than that
   of C02 but consistent within $1\sigma$.

5. The amplitude of the power spectrum rises with increasing redshift 
   due to the increasing mean flux. After correcting to the same mean
   flux the fluctuation amplitude decreases with increasing
   redshift as expected from gravitational clustering.

\section*{Acknowledgments.} 
This work was supported by the European Community Research and
Training Network ``The Physics of the Intergalactic Medium''.
We would like to thank ESO, the ESO staff, the ESO science 
verification team  and the team of the UVESLP 
``QSO absorption lines'' for initiating, 
compiling and making publicly available  a superb set of QSO
absorption spectra. Last but not least we would like to thank the UVES 
team for building the spectrograph. TSK would like to thank
Michael Rauch for his advice on  continuum fitting. We thank the
anonymous referee for his very helpful report and D. Weinberg for
useful discussions.

\begin{table}
\caption{Mean 1D  flux power spectrum of the full sample
($<z>=2.25$) for different flux estimators $^{a}$}
\label{tab3}
\begin{tabular}{lccc}
\hline
\noalign{\smallskip}
k (s/km) & $P_{F1}(k)$ (km/s) & $P_{F2}(k)$ (km/s) & $P_{F3}(k)$ (km/s)\\
\noalign{\smallskip}
\hline
\noalign{\smallskip}
 0.0010  & 13.0770 $\pm$  2.1244 & 25.9785 $\pm$  6.6807 & 47.6964 $\pm$ 16.1311	\\
 0.0013  & 13.7172 $\pm$  1.5985 & 25.8181 $\pm$  5.3323 & 35.9422 $\pm$  5.9444 \\
 0.0018  & 15.7603 $\pm$  2.1502 & 29.7464 $\pm$  6.0253 & 35.8427 $\pm$  6.9350 \\
 0.0024  & 14.5357 $\pm$  1.6421 & 25.8690 $\pm$  4.2306 & 33.7291 $\pm$  6.7301 \\
 0.0032  & 14.8066 $\pm$  1.3860 & 27.9988 $\pm$  5.5553 & 32.3716 $\pm$  6.6544 \\
 0.0042  & 12.4382 $\pm$  0.7775 & 23.0850 $\pm$  4.1762 & 25.3903 $\pm$  6.5073 \\
 0.0056  & 11.5405 $\pm$  1.0521 & 22.2407 $\pm$  4.5617 & 23.3662 $\pm$  6.6347 \\
 0.0075  &  9.8115 $\pm$  0.6149 & 18.3001 $\pm$  2.7549 & 19.4855 $\pm$  3.5507 \\
 0.0100  &  7.7051 $\pm$  0.4625 & 14.7067 $\pm$  2.6998 & 15.0972 $\pm$  2.7572 \\
 0.0133  &  7.0616 $\pm$  0.4988 & 13.2415 $\pm$  2.3227 & 12.9928 $\pm$  2.5944 \\
 0.0178  &  4.9893 $\pm$  0.3464 &  9.7117 $\pm$  1.9065 & 10.2946 $\pm$  2.2990 \\
 0.0237  &  3.4304 $\pm$  0.2170 &  7.0356 $\pm$  1.7857 &  7.0597 $\pm$  1.8803 \\
 0.0316  &  2.1377 $\pm$  0.1390 &  4.3820 $\pm$  1.1375 &  4.5441 $\pm$  1.3845 \\
 0.0422  &  1.2155 $\pm$  0.0786 &  2.4105 $\pm$  0.5488 &  2.3807 $\pm$  0.6219 \\
 0.0562  &  0.6369 $\pm$  0.0483 &  1.3610 $\pm$  0.4058 &  1.4009 $\pm$  0.5125 \\
 0.0750  &  0.2686 $\pm$  0.0180 &  0.5670 $\pm$  0.1649 &  0.5864 $\pm$  0.2019 \\
 0.1000  &  0.1008 $\pm$  0.0074 &  0.2230 $\pm$  0.0752 &  0.2403 $\pm$  0.0890 \\
 0.1334  &  0.0389 $\pm$  0.0030 &  0.0817 $\pm$  0.0245 &  0.0889 $\pm$  0.0286 \\
 0.1778  &  0.0157 $\pm$  0.0011 &  0.0309 $\pm$  0.0075 &  0.0355 $\pm$  0.0087 \\
 0.2371  &  0.0071 $\pm$  0.0006 &  0.0134 $\pm$  0.0026 &  0.0158 $\pm$  0.0027 \\
\noalign{\smallskip}
\hline
\end{tabular}
\begin{list}{}{}
\item[a] The different flux estimators: $F1=\exp(-\tau)$; $F2=
e^{-\tau}/<e^{-\tau}>-1$, with $<e^{-\tau}>$ the mean flux level for
each QSOs; $F3_{s25} = I_{\rm obs}/<I_{\rm obs}>_{s25}-1$.  Note that
$F1$ and $F2$ are computed from continuum fitted spectra, while
$F3_{s25}$ is calculated from not continuum fitted spectra smoothed
with a Gaussian window of 25 \AA \, width.  See Section 3.3 for more
details. Note that this table is different from the one in the
published version in Kim et al. (2004) MNRAS 347, 355 (where the $k$
values of the 1D flux power spectrum had been erroneously shifted by
half a bin size in $\log k$).  A machine-readable version of this table
is available at {\tt http://www.ast.cam.ac.uk/$\sim$rtnigm/luqas.htm}.
\end{list}
\end{table}

\begin{table}
\caption{
Mean 3D  flux power spectrum of the full sample ($<z>=2.25$) 
for different flux estimators $^{a}$}
\label{tab4}
\begin{tabular}{lccc}
\hline
\noalign{\smallskip}
k (s/km) & $\Delta^2_{F1}(k)$  & $\Delta^2_{F2}(k)$ &
$\Delta^2_{F3}(k)$\\
\noalign{\smallskip}
\hline
\noalign{\smallskip}
 0.0036 &  0.0154 $\pm$  0.0108  &  0.0236 $\pm$  0.0183  & 0.0381 $\pm$  0.0294\\
 0.0046 &  0.0044 $\pm$  0.0064  &  0.0103 $\pm$  0.0122	&  0.0105 $\pm$  0.0183\\
 0.0060 &  0.0074 $\pm$  0.0084  &  0.0061 $\pm$  0.0148	&  0.0179 $\pm$  0.0213\\
 0.0077 &  0.0181 $\pm$  0.0101  &  0.0403 $\pm$  0.0192	&  0.0403 $\pm$  0.0357\\
 0.0100 &  0.0116 $\pm$  0.0109  &  0.0215 $\pm$  0.0238	&  0.0432 $\pm$  0.0312\\
 0.0129 &  0.0237 $\pm$  0.0116  &  0.0430 $\pm$  0.0196	&  0.0424 $\pm$  0.0267\\
 0.0167 &  0.0498 $\pm$  0.0134  &  0.0756 $\pm$  0.0242	&  0.0871 $\pm$  0.0245\\
 0.0215 &  0.0347 $\pm$  0.0094  &  0.0776 $\pm$  0.0340	&  0.0813 $\pm$  0.0283\\
 0.0278 &  0.0527 $\pm$  0.0070  &  0.0989 $\pm$  0.0193	&  0.0981 $\pm$  0.0195\\
 0.0359 &  0.0464 $\pm$  0.0052  &  0.0992 $\pm$  0.0335	&  0.1179 $\pm$  0.0436\\
 0.0464 &  0.0409 $\pm$  0.0068  &  0.0711 $\pm$  0.0146	&  0.0675 $\pm$  0.0138\\
 0.0599 &  0.0416 $\pm$  0.0053  &  0.0906 $\pm$  0.0238	&  0.0978 $\pm$  0.0279\\
 0.0774 &  0.0272 $\pm$  0.0025  &  0.0531 $\pm$  0.0153	&  0.0577 $\pm$  0.0167\\
 0.1000 &  0.0189 $\pm$  0.0021  &  0.0445 $\pm$  0.0155	&  0.0469 $\pm$  0.0180\\
 0.1292 &  0.0082 $\pm$  0.0007  &  0.0174 $\pm$  0.0059	&  0.0188 $\pm$  0.0054\\
 0.1668 &  0.0048 $\pm$  0.0006  &  0.0106 $\pm$  0.0039	&  0.0116 $\pm$  0.0048\\
 0.2154 &  0.0023 $\pm$  0.0002  &  0.0047 $\pm$  0.0013	&  0.0051 $\pm$  0.0014\\
\noalign{\smallskip}
\hline
\end{tabular}
\begin{list}{}{}
\item[a]
The different flux estimators: $F1=\exp(-\tau)$, 
$F2= e^{-\tau}/<e^{-\tau}>-1$,  $F3_{s25} = I_{\rm obs}/<I_{\rm obs}>_{s25}-1$. 
Note that $F1$ and $F2$ are computed from continuum fitted spectra, while
$F3_{s25}$  is calculated from not continuum fitted spectra smoothed with a 
Gaussian window of 25 \AA\, width. 
See Section 3.3 for
more details.
A machine-readable version of this table is available at {\tt http://www.ast.cam.ac.uk/$\sim$rtnigm/luqas.htm}.
\end{list}
\end{table}

\begin{table}
\caption{1D Flux power spectrum for $F2=\exp(-\tau)/<\exp(-\tau)>-1$ in three different
redshift ranges $^{a}$ }
\label{tab5}
\begin{tabular}{lccc}
\hline
\noalign{\smallskip}
k (s/km) & $P_{F2}(k) \,({\rm km/s}) \;\; <z>=1.87$  & $P_{F2}(k) \,({\rm km/s}) \;\; <z>=2.18$ &
$P_{F2}(k) ({\rm km/s}) \;\; <z>=2.58$\\
\noalign{\smallskip}
\hline
\noalign{\smallskip}
 0.0010 & 26.3677 $\pm$  8.3099  & 14.2952 $\pm$  4.0715  &  9.9894 $\pm$  3.3723\\
 0.0013 & 26.2042 $\pm$  7.0032  & 22.3200 $\pm$ 11.3997	 & 33.3230 $\pm$ 10.7060\\
 0.0018 & 18.0914 $\pm$  4.9298  & 16.8461 $\pm$  2.4814	 & 46.9632 $\pm$ 19.5921\\
 0.0024 & 19.2437 $\pm$  6.1694  & 15.5272 $\pm$  2.4619	 & 23.3280 $\pm$  5.3857\\
 0.0032 & 18.6062 $\pm$  4.0084  & 18.5186 $\pm$  4.0294	 & 33.7688 $\pm$ 10.1441\\
 0.0042 & 13.8969 $\pm$  2.2267  & 17.0164 $\pm$  3.2064	 & 21.2278 $\pm$  2.6202\\
 0.0056 & 13.0126 $\pm$  1.4421  & 14.0088 $\pm$  1.6224	 & 23.6721 $\pm$  4.1560\\
 0.0075 & 11.1447 $\pm$  1.7044  & 12.9511 $\pm$  2.4550	 & 21.2667 $\pm$  4.3868\\
 0.0100 &  9.7740 $\pm$  1.1599  &  9.2170 $\pm$  0.9704	 & 15.7910 $\pm$  2.4561\\
 0.0133 &  6.5832 $\pm$  0.5700  &  9.1975 $\pm$  0.8948	 & 13.5331 $\pm$  1.5505\\
 0.0178 &  4.8737 $\pm$  0.5387  &  6.4600 $\pm$  0.4893	 & 10.3522 $\pm$  1.1515\\
 0.0237 &  3.3942 $\pm$  0.3616  &  4.2577 $\pm$  0.3428	 &  7.8964 $\pm$  0.9131\\
 0.0316 &  2.3183 $\pm$  0.1674  &  2.4256 $\pm$  0.1518	 &  4.4644 $\pm$  0.3660\\
 0.0422 &  1.1390 $\pm$  0.1046  &  1.4664 $\pm$  0.1074	 &  2.7588 $\pm$  0.2459\\
 0.0562 &  0.6504 $\pm$  0.0399  &  0.7202 $\pm$  0.0418	 &  1.2996 $\pm$  0.0980\\
 0.0750 &  0.2855 $\pm$  0.0214  &  0.3304 $\pm$  0.0244	 &  0.5284 $\pm$  0.0499\\
 0.1000 &  0.1197 $\pm$  0.0097  &  0.1213 $\pm$  0.0094	 &  0.2002 $\pm$  0.0164\\
 0.1334 &  0.0535 $\pm$  0.0042  &  0.0490 $\pm$  0.0060	 &  0.0827 $\pm$  0.0176\\
 0.1778 &  0.0241 $\pm$  0.0026  &  0.0199 $\pm$  0.0022	 &  0.0335 $\pm$  0.0062\\
 0.2371 &  0.0117 $\pm$  0.0011  &  0.0089 $\pm$  0.0009	 &  0.0143 $\pm$  0.0029\\
\noalign{\smallskip}
\hline
\end{tabular}
\begin{list}{}{}
\item[a]
Note that this table is different from the one in the
published version in Kim et al. (2004) MNRAS 347, 355 (where the $k$
values of the 1D flux power spectrum had been erroneously shifted by
half a bin size in $\log k$). A machine-readable version of this table is available at {\tt http://www.ast.cam.ac.uk/$\sim$rtnigm/luqas.htm}.
\end{list}
\end{table}

\begin{table}
\caption{``3D'' Flux power spectrum for $F2=\exp(-\tau)/<\exp(-\tau)>-1$ in three different
redshift ranges  $^{a}$}
\label{tab6}
\begin{tabular}{lccc}
\hline
\noalign{\smallskip}
k (s/km) & $\Delta^2_{F2}(k)\;\; <z>=1.87$  & $\Delta^2_{F2}(k)\;\; <z>=2.18$ &
$\Delta^2_{F2}(k)\;\; <z>=2.58$\\
\noalign{\smallskip}
\hline
\noalign{\smallskip}
 0.0048 &  0.0118 $\pm$  0.0134  &  0.0152 $\pm$  0.0255 &  0.0187
$\pm$  0.0167 \\
 0.0069 &  0.0137 $\pm$  0.0163  &  0.0118 $\pm$  0.0106 &  0.0134 $\pm$  0.0288\\
 0.0099 &  0.0218 $\pm$  0.0177  &  0.0387 $\pm$  0.0151 &  0.0387 $\pm$  0.0445\\
 0.0143 &  0.0445 $\pm$  0.0153  &  0.0200 $\pm$  0.0125 &  0.0864 $\pm$  0.0372\\
 0.0205 &  0.0456 $\pm$  0.0131  &  0.0721 $\pm$  0.0133 &  0.0620 $\pm$  0.0208\\
 0.0296 &  0.0474 $\pm$  0.0059  &  0.0680 $\pm$  0.0075 &  0.1180 $\pm$  0.0208\\
 0.0426 &  0.0509 $\pm$  0.0041  &  0.0605 $\pm$  0.0069 &  0.1117 $\pm$  0.0162\\
 0.0613 &  0.0444 $\pm$  0.0040  &  0.0455 $\pm$  0.0035 &  0.1048 $\pm$  0.0106\\
 0.0883 &  0.0248 $\pm$  0.0026  &  0.0335 $\pm$  0.0031 &  0.0526 $\pm$  0.0058\\
 0.1271 &  0.0119 $\pm$  0.0015  &  0.0136 $\pm$  0.0015 &  0.0183 $\pm$  0.0020\\
 0.1830 &  0.0053 $\pm$  0.0008  &  0.0054 $\pm$  0.0008 &  0.0107 $\pm$  0.0039 \\
\noalign{\smallskip}
\hline
\end{tabular}
\begin{list}{}{}
\item[a]
A machine-readable version of this table is available at {\tt http://www.ast.cam.ac.uk/$\sim$rtnigm/luqas.htm}.
\end{list}
\end{table}

\end{document}